\documentclass[epjST]{svjour}

\usepackage{graphics}
\usepackage{amssymb,amsfonts,amsmath}
\usepackage{graphicx}
\usepackage{color}

\newcommand{\B}{{\textsc{b}}}

\begin{document}

\title{Universality at work --- the local sine-Gordon model, lattice fermions,
  and quantum circuits}

\author{A.~Anthore\inst{1,2} \and D.M.~Kennes\inst{3} \and
  E.~Boulat\inst{4} \and S.~Andergassen\inst{5} \and F.~Pierre\inst{1} \and
  V.~Meden\inst{3}\fnmsep\thanks{\email{meden@physik.rwth-aachen.de}}}

\institute{Centre de Nanosciences et de Nanotechnologies (C2N), CNRS,
  Univ Paris Sud, Universit\'e Paris-Saclay, 91120 Palaiseau, France
  \and Universit\'e de Paris, Univ Paris Diderot, 75013 Paris, France
  \and Institut f\"ur Theorie der Statistischen Physik, RWTH Aachen University
  and JARA-Fundamentals of Future Information Technology,
  52056 Aachen, Germany
  \and Universit\'e de Paris, Laboratoire Mat\'eriaux et Ph\'enom\`enes Quantiques (MPQ),
  Univ Paris Diderot, CNRS, 75013 Paris, France
  \and Institut f\"ur Theoretische Physik and Center for Quantum Science,
  Universit\"at T\"ubingen, 72076 T\"ubingen, Germany}

  \abstract{
  We review the intriguing many-body physics resulting out of the interplay of a single,
  local impurity and the two-particle interaction in a one-dimensional
  Fermi system. Even if the underlying homogeneous correlated
  system is taken to be metallic, this interplay leads to an emergent
  quantum phase transition
  between metallic and insulating states. We show that the zero temperature
  critical point and the universal low-energy physics associated to it, is realized
  in two different models, the field theoretical local sine-Gordon model and
  spinless fermions on a lattice with nearest-neighbor hopping and two-particle
  interaction, as well as in an experimental setup consisting of a highly tunable
  quantum circuit. Despite the different high-energy physics of the three systems
  the universal low-energy scaling curves of the conductance as a function of
  temperature agree up to a very high precision without
  any free parameter. Overall this provides a convincing example of how emergent
  universality in complex systems originating from a common underlying quantum critical
  point establishes a bridge between different fields of physics. In our case between
  field theory, quantum many-body theory of correlated
  Fermi systems, and experimental circuit quantum electrodynamics.   
}

\maketitle

\section{An impurity in a one-dimensional, correlated Fermi system}
\label{sec:impin1d}

\subsection{Linear response theory}
\label{subsec:linresp}

Linear response theory provides a framework to study the effect of a weak
local impurity in a nonrelativistic many-body Fermi system. Let us assume that
we are interested in
the change of the particle density in response to the introduction of an impurity
potential into a translational invariant system. In this case we need to consider the
density-density response function $\chi(q,\omega)$ of the homogeneous system. In one
spatial dimension (1d) and in the absence of two-particle interaction this Lindhard
function at temperature $T=0$ and in the static limit, i.e.~at energy $\omega=0$
(measured relative to the chemical potential), shows a logarithmic divergence
if the momentum $q$ approaches twice the Fermi momentum $k_{\rm F}$ \cite{Fetter71}.
This divergence results from the restricted phase space available in 1d and is thus not specific
to any particular model such as, e.g., the 1d Fermi gas. The divergence indicates
that any impurity with a nonvanishing backscattering $2 k_{\rm F}$ component strongly
affects the density. Taking into account that in 1d particles cannot bypass the impurity
this insight might be considered as intuitive. 

In 1974 the density-density response function of the spinless Fermi gas complemented
by a two-particle interaction was computed \cite{Luther74,Mattis74}. It was shown
that the logarithmic singularity of the noninteracting case turns into a power law
with an exponent which can be expressed in terms of an interaction dependent
parameter $K$, the so-called Tomonaga-Luttinger liquid parameter (see below),
\begin{equation}
  \chi(q \approx 2 k_{\rm F},0) \sim |q-2 k_{\rm F}|^{2(K-1)} .
\label{eq:chi}
\end{equation}  
For repulsive two-particle interactions $0< K < 1$ holds while $K>1$ for attractive ones.
The power-law divergence in the repulsive case shows that the interacting homogeneous system
is perturbed even more strongly by a single impurity than the noninteracting one.
It, in fact, indicates that linear response theory breaks down. The response of the
homogeneous system to the impurity potential is large even if its amplitude at
momentum transfer $2 k_{\rm F}$ is arbitrarily small. Thus more elaborate
methods than linear response theory are required to study the effect of impurity backscattering in
1d correlated Fermi systems. Before returning to this issue in Sect.~\ref{subsec:impback}
we will next discuss
the physics of homogeneous interacting 1d Fermi systems from a more general
perspective. This turns out to be a necessary first step.

\subsection{Homogeneous Tomonaga-Luttinger liquids}
\label{subsec:homTL}

Power-law scaling of correlation functions as in Eq.~(\ref{eq:chi}) is
characteristic for translationally invariant, metallic, and interacting 1d Fermi
systems. Employing renormalization group (RG) arguments, one can show that the
Tomonaga-Luttinger model is the low-energy fixed point of a large class
of models in which the interaction does not lead to the opening of a
gap \cite{Solyom79,Giamarchi03}. This class includes the continuum
electron gas with two-particle interaction but also lattice models, such as,
e.g., spinless fermions with nearest-neighbor hopping and nearest-neighbor
interaction. The Tomonaga-Luttinger model plays the same role
in 1d as the noninteracting Fermi gas does for (interacting) higher
dimensional metallic systems. The noninteracting Fermi gas is the
fixed-point model of systems falling into the Fermi liquid 
universality class.

To understand the universal
low-energy physics of Tomonaga-Luttinger liquids, in a first step, one can thus
study the Tomonaga-Luttinger model. It has two branches of fermions
(right- and left-moving ones) with linear dispersion and two-particle scattering
which is restricted to small momentum transfer $|q| \ll k_{\rm F}$. The
elementary low-energy excitations of this model are not given by fermionic
quasi-particles, as in Fermi liquids, but are of collective bosonic nature. Using
bosonization \cite{Giamarchi03,vonDelft98,Schoenhammer05} thermodynamic observables
and all correlation functions of interest can be computed exactly in the
low-energy scaling limit. For the spinless case, on which we focus, the low-energy
physics is characterized by only two parameters, $K$ and the renormalized
velocity $v$. The parameters appearing in the Hamiltonian of the Tomonaga-Luttinger
model enter the exponents of the characteristic power laws of correlation functions
only via $K$. The exponent of each correlation function is given by a
unique function of $K$. For an example, see Eq.~(\ref{eq:chi}).

To determine the low-energy physics of a given microscopic model from the
Tomo\-naga-Luttinger liquid
universality class one can, in a second step, proceed as follows. By computing two
thermodynamic observables, e.g., the compressibility and the specific heat,
$K$ and $v$ can be determined in terms of the microscopic parameters. In particular,
these are the amplitude and range of the two-particle interaction, the parameters of the
single-particle dispersion (such as the hopping amplitude in a tight-binding model),
and the band filling. $K$ and $v$ can then be plugged into the expressions of
the correlation functions and observables of the Tomonaga-Luttinger model.
The challenging explicit computation of correlation functions for an interacting
microscopic model of interest can be avoided this way. Thermodynamic observables
are easier to access either by analytical means such as 
perturbation theory and the Bethe ansatz, or by numerical
approaches \cite{Giamarchi03,Schoenhammer05}.

For the above mentioned lattice model of spinless fermions, which will be one of the
models considered here, $K$ and $v$ as functions of the nearest-neighbor interaction
$U$, the nearest-neighbor hopping $t$, and the filling $\nu$ can be computed
exactly (see below) employing the Bethe ansatz expression for the ground state
energy \cite{Haldane80}. At half filling $\nu=1/2$ the corresponding set of integral
equations can be solved analytically while away from half filling a numerical solution
up to very high precision is possible. For $\nu \neq 1/2$ the model is from the
Tomonaga-Luttinger liquid universality class for all $U/t>-2$, while for $\nu = 1/2$ this
low-energy physics is only found for $-2 < U/t <2$. 
In Sect.~\ref{sec:lattice_fermions} we will return to this model.

For many years emergent universal Tomonaga-Luttinger liquid physics of 1d systems
was considered to be an appealing theoretical concept, which, however, was far from
being realizable in real-world experiments. Only at the beginning of the 1990s
material science and nanostructuring techniques reached a level, such that an experimental
realization appeared to be within reach. Promising systems to observe the typical
Tomonaga-Luttinger liquid power-law behavior of the spectral and transport properties
are highly anisotropic quasi 1d crystals, semi-conductor-based quasi 1d
heterostructures (cleaved-edge overgrowth), self-organized atom chains on surfaces,
and unidirectional long molecules, such as, e.g., metallic carbon nanotubes. In fact,
examples from all these classes were investigated concerning their Tomonaga-Luttinger liquid
properties (for reviews, see Ref.~\cite{Grioni09,Deshpande10,Giamarchi12}). Although many
of the measurements are consistent with Tomonaga-Luttinger liquid behavior,
convincing examples of power-law scaling in an energy variable, such as 
the temperature $T$ or the frequency $\omega$, are rare. In most experiments the energy
regime over which results consistent with power-law behavior can be observed is small,
typically less than one order of magnitude. At the lower end the power laws are cut off
by a finite energy resolution or effects beyond the Tomonaga-Luttinger liquid theory
such as coupling of the 1d chains. At the higher end details of the experimental
system, e.g. the band structure, cut off the universal low-energy physics.
Having only a small window of energies available and taking into account experimental
noise, it is hardly possible to distinguish power-law behavior from other functional
forms. In addition, in most systems, it is impossible to control any other parameter
besides the energy variable in which power-law scaling is investigated, e.g. the
strength of the two-particle interaction. It is thus impossible to show consistency
of the experimentally extracted exponents with the predictions of
Tomonaga-Luttinger liquid theory. In short, the systems lack control and tunability.
This calls for further attempts to realize Tomonaga-Luttinger liquid physics in
experiments. One way is to use other systems, e.g. quantum circuits, to emulate
Tomonaga-Luttinger liquid behavior (see, e.g. Ref.~\cite{Safi04}). Here we will follow
this route but consider inhomogeneous Tomonaga-Luttinger liquids instead of translational
invariant ones.  

\subsection{Impurity backscattering}
\label{subsec:impback}

In a first step towards a comprehensive understanding of the effect of a localized
impurity in a Tomonaga-Luttinger liquid beyond linear response theory (see Sect.~\ref{subsec:linresp}),
in 1982 Apel and Rice studied a system in which
the Tomonaga-Luttinger model is complemented by a pure single-particle
backscattering term \cite{Apel82}. Rewritten in terms of the bosonic fields
of the bosonization approach it becomes obvious that at low energies this model is equal to
the local sine-Gordon model studied in quantum field theory \cite{Giamarchi03,vonDelft98,Schoenhammer05}.
Apel and Rice used a RG-related scaling theory. Their results indicate that
for a repulsive two-particle interaction with $0< K < 1$ even a weak impurity drives
the system from being a metal with finite (dc) conductance $G$ at vanishing
$T$ into an insulating phase with $G(T=0)=0$. For attractive
interactions the system, in contrast, remains  metallic.

In 1992 this picture was confirmed by Kane and Fisher \cite{Kane92}. They studied
the local sine-Gordon model for arbitrary $K$ using a RG approach
perturbative in the backscattering amplitude. This calculation was complemented by a
perturbative RG for the (dual, see Ref.~\cite{Kane92}) problem of a weak link
connecting two semi-infinite wires each modeled by the Tomonaga-Luttinger
Hamiltonian. It turned out that the perfect chain fixed point with
vanishing impurity backward scattering is stable for attractive two-particle
interactions with $K>1$, but unstable for repulsive ones with $0<K<1$,
while the cut chain fixed point, with $G=0$,  is stable in the repulsive case and unstable
for attractive interactions. For $K=1/2$ it was possible to show that both fixed
points are directly linked, i.e. not separated by any intermediate fixed point.
The consequences of this behavior for observables is best illustrated considering
the temperature dependence of the linear conductance. We will discuss this in the next
subsection. We note that in their analysis Kane and Fisher were able to resort to
results obtained for Hamiltonians studied earlier in the field of dissipative
quantum systems \cite{Chakravarty82,Schmid83}. 

The absence of any intermediate fixed point in the local
sine-Gordon model with $K=1/n$, $n \in \mathbb{N}$,
was confirmed numerically by quantum
Monte Carlo approaches \cite{Moon93,Leung95} as well as analytically by the Bethe
ansatz solution \cite{Fendley95}. This let to the general expectation that
the same will hold for the local sine-Gordon model with arbitrary $K$. Very
recently a modified Bethe ansatz was used to solve the local sine-Gordon model
first for $K=2/3$ \cite{Anthore18} and later for all rational $K<1$ \cite{Boulat19}.
We will elaborate on these solutions in Sect.~\ref{sec:sine_Gordon}.

The just described physics of the local sine-Gordon model is often phrased as
follows. For repulsive interactions and on small energy scales even a weak
impurity grows and effectively cuts the chain into two parts. The opposite
holds for attractive interactions. Even starting with only a weak link
connecting two semi-infinite chains (i.e. a strong impurity) the system
is ``healed'', that is, the impurity vanishes.

In the early to mid 1990s two important steps were taken, to
show that this intriguing many-particle physics, resulting out of
the interplay of single-particle backscattering and the two-particle interaction,
is also realized in other models for a single impurity in a 1d correlated
fermion system than the specific local sine-Gordon model. Numerical results
for the above mentioned lattice model of spinless fermions (more precisely, for
the equivalent XXZ-Heisenberg model) at small system sizes turned out to be
consistent with the above RG flow in the two limits of a weak impurity and a
weak link \cite{Eggert92}. In addition, a fermionic RG applicable in the limit
$K \to 1$ was set up for the 1d continuum electron gas with an impurity and
showed the RG flow connecting the two fixed points \cite{Yue94}. However, it took another
ten years to develop an approximate RG method which is capable to capture
the full crossover from the perfect to the cut chain fixed point
(or vice versa) for a lattice model \cite{Meden08}.
In Sect.~\ref{sec:lattice_fermions} we will describe this
approach and present results for $G(T)$ obtained this way. 

Based on these insights one can now be certain that the local sine-Gordon model
is the effective low-energy model of a large class of impurity Hamiltonians
with the bulk part falling into the Tomonaga-Luttinger liquid universality class.
To avoid any confusion we emphasize that in microscopic models one has to
distinguish two RG flows. The one of the bulk part of a given model towards the
Tomonaga-Luttinger model and the flow of the single-particle (impurity) backscattering
amplitude. 

The above described transition from metallic $G(T=0) >0 $ to insulating $G(T=0)=0$
behavior can also be understood within the framework of quantum critical behavior \cite{Sachdev11}.
The Tomonaga-Luttinger liquid parameter $K$ can be used to tune the inhomogeneous
system through the $T=0$ quantum phase transition from a metal $K > 1$ to an
insulator $0<K<1$. The temperature dependence of the conductance, to be discussed
in the next section, reflects the scaling away from criticality which is dominated
by the quantum critical point. This quantum critical perspective on the problem
of a single impurity in a Tomonaga-Luttinger liquid helps to understand
the universality across models and experimental systems; see below.

\subsection{The linear conductance and the universal $\boldsymbol{\beta}$-function}
\label{subsec:lincond}

Within the local sine-Gordon model employing the RG approach perturbative
in the weak impurity \cite{Kane92} it is straightforward to show that the conductance close to the
perfect chain fixed point depends on temperature as
\begin{equation}
G_0 - G(T) \sim V_{\rm b}^2 T^{2(K-1)} ,
  \label{eq:Gweakimp}
\end{equation}  
with $G_0$ being the $T=0$ conductance in the absence of the impurity (see below) and
$V_{\rm b}$ the bare (as opposed to renormalized) amplitude of the impurity
backscattering. For $K>1$ the exponent is
positive and the finite temperature correction vanishes. This indicates the stability
of the perfect chain fixed point for attractive interactions. For $0 < K <1$,
the correction grows. One says that the backscattering renormalizes towards strong
coupling. As the calculation leading to Eq.~(\ref{eq:Gweakimp}) is controlled
for small renormalized backscattering it holds as long as the right hand side
remains small. Assuming a fixed $V_{\rm b}$ this is only the case for not too small
temperatures. 

A similar analysis close to the cut chain fixed point gives
\begin{equation}
G(T) \sim t_{\rm wl}^2 T^{2(1/K-1)} ,
  \label{eq:Gweaklink}
\end{equation}  
where $t_{\rm wl}$ is a measure for the bare hopping between the two semi-infinite
chains. For $0<K<1$ the conductance vanishes for $T \to 0$; the cut chain fixed point
is stable and the system is an insulator. For $K>1$ the finite $T$ correction to the
cut chain conductance $G=0$ grows for decreasing $T$. As the result
Eq.~(\ref{eq:Gweaklink}) was derived in the weak link limit the right hand side
has to stay small and the temperature cannot be taken too small. In other words, the
cut chain fixed point is unstable. 

As no intermediate fixed point interrupts the RG flow from the perfect to the cut chain fixed
point (or vice versa), the limiting behavior of Eqs.~(\ref{eq:Gweakimp}) and (\ref{eq:Gweaklink})
are connected by a unique, $K$-dependent conductance function. 
More generally, such universal scaling behavior as a function of temperature is expected in the
vicinity of continuous quantum phase transitions, when the system is slightly detuned from the quantum
critical point \cite{Sachdev11}. The theory of quantum critical phenomena predicts that all microscopic
parameters [e.g.~$V_{\rm b}$ of Eq.~(\ref{eq:Gweakimp}) and $t_{\rm wl}$ of Eq.~(\ref{eq:Gweaklink})]
can be encapsulated into a rescaling temperature $T_0$, such that the conductance is given by
$G_0 G_{K}(T/T_0)$ with a universal---but $K$-dependent---dimensionless function $G_{K}(x)$ and $x=T/T_0$.

However, the numerical value of $T_0$ is a priori unknown, and depends
on specific details of the system. In order to perform a direct comparison between different models, in
our case the local sine-Gordon model and the lattice model of spinless fermions, as well as a
comparison to experiments, it is possible to eliminate $T_0$ by considering, instead of $G$, its
logarithmic derivative $dG/d\ln T$ that does not depend on the temperature scale. An underlying universal
scaling law then implies the existence of a so-called $\beta$-function completely characterizing the
conductance renormalization flow through the relation
\begin{equation}
\beta_K = \frac{dg(T)}{d \ln (T)}, 
  \label{eq:defbeta}
\end{equation}  
with the normalized conductance $g(T) = \frac{G(T)}{G_0}$.

In the two limits of Eqs.~(\ref{eq:Gweakimp}) and (\ref{eq:Gweaklink}) we obtain
\begin{equation}
\beta_K = 2(1-K)(1-g)
  \label{eq:betaweakimp}
\end{equation}
for $1-g \ll 1$ and
\begin{equation}
\beta_K  =  2(1/K-1) g 
  \label{eq:betaweaklink}
\end{equation}  
for $g \ll 1$. In these limits $\beta_K$ depends on $x=T/T_0$ only via $g$.
Based on the discussion of Sect.~\ref{subsec:impback} we expect that 
$\beta_K$ is also model independent in the above two limits.

From the exact analytical results for $G(T)$ obtained by Bethe ansatz for the local sine-Gordon
model---for details see the next section---it is known that $\beta_K$ is a function of
$g$ only also for arbitrary $g$, not only in the limits $1-g \ll 1$ and $g \ll 1$.
If the idea of quantum critical universality holds and taken that the quantum critical
point separating the metal from the insulator is the same in all models of
Tomonaga-Luttinger liquids with impurity backscattering, the same function $\beta_K(g)$
should be found in microscopic models for the specific value of $K$ of the
underlying homogeneous system.

\begin{figure}
  \begin{center}
  \includegraphics[width=0.4\textwidth,clip]{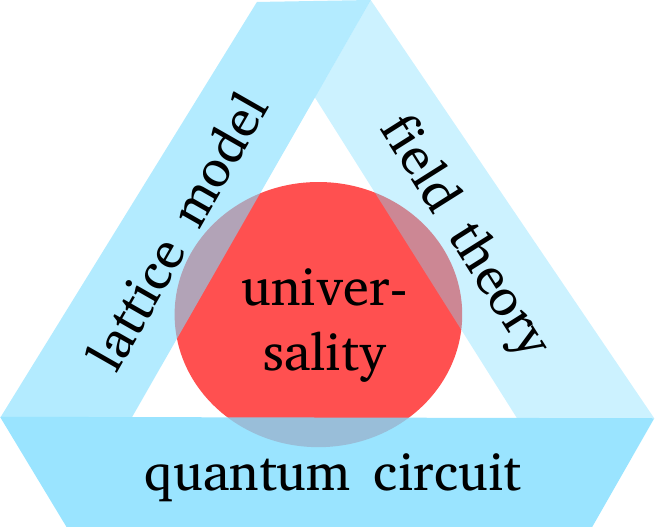}
  \caption{A sketch of our ``triangle of universality''. The two models and the experimental system
    show the same universal low-energy physics resulting out of a common underlying quantum critical
    point.} 
  \label{fig0}
  \end{center}
\end{figure}

It is exactly this type of emergent universality which
we will demonstrate in the remainder
of this minireview. We will explicitly verify that the $\beta$-function of the 
lattice model of spinless fermions with nearest-neighbor hopping,
nearest-neighbor interaction, and an impurity falls on top of the $\beta$-function
of the local sine-Gordon model given the same $K$ (but without any free parameter).
Furthermore, we will show that the measured conductance of a highly tunable quantum
circuit \cite{Anthore18}, described by a Hamiltonian which can be argued to be equivalent to 
the local sine-Gordon model at low energy scales, leads to a $\beta$-function
also falling on the universal curve. This will complete our ``triangle of universality'';
see Fig.~\ref{fig0} for an illustration.
We note that related experimental
results were obtained earlier in Ref.~\cite{Mebrahtu12}, but for a circuit setup which was
less tunable and provided access to a single value of $K$ only.
Only recently is was possible to close the triangle with the sides being a
field theory (the local sine-Gordon model), an interacting, microscopic lattice model, and an experimental
quantum circuit for different values of $K$. It exploits the latest
technical developments in the Bethe ansatz solution of the local sine-Gordon model \cite{Boulat19}
and the recently achieved tunability of $K$ in the quantum circuits \cite{Anthore18}. In both
research fields only now values of $K$ sufficiently close to 1 can be reached, for which controlled
results for the $\beta$-function of the interacting fermionic system are
available \cite{Meden08}. 

At the same time, the quantum circuit provides a convincing experimental emulation of
(inhomogeneous) Tomonaga-Luttinger liquid physics beyond the limitations
discussed in Sect.~\ref{subsec:homTL}.

We next review how to obtain $G_{K}(T/T_0)$ or, equivalently, $\beta_K(g)$ for the
local sine-Gordon model.

\section{The local sine-Gordon model}
\label{sec:sine_Gordon}

The local sine-Gordon model is a minimal quantum impurity model, of a free 1d massless chiral
(say right-moving) boson described by the field $\phi(x)$, with $x \in {\mathbb R}$, which, at the origin, is
perturbed by a cosine potential of amplitude $\gamma$ 
\begin{equation}
H=H_0[\phi]+H_\B ,  \quad H_0[\phi] = \int_{-\infty}^{\infty}dx\,\big(\partial_x\phi\big)^2, \quad
H_\B = \gamma\cos\left[ \beta\phi(0)\right] . 
\label{HBSG}
\end{equation}
We first discuss how it is related to the problem of a single impurity in a 1d
fermionic many-body system and, secondly, review its exact solution by means of a Bethe
ansatz approach. We consider both, the so-called diagonal case, when
$\lambda = \frac{8\pi}{\beta^2} -1 \in {\mathbb N}$ \cite{Fendley95},
and, the more general off-diagonal case, when $\lambda\in\mathbb{Q}^+$
\cite{Boulat19}.
We discuss the current $I$ as a function of $T$ but also as a function of a bias
voltage $V$ applied across the impurity (out-of-equilibrium setup). From this
the linear conductance $G(T)= \left. \frac{\partial I}{\partial V}\right|_{V=0}$
can be computed. 

Employing bosonization the low-energy features of the Tomonaga-Luttinger liquid model are 
captured by a 1d free massless boson theory \cite{Giamarchi03,vonDelft98,Schoenhammer05},
that can be decomposed in terms
of chiral  right-moving  and left-moving modes, with Hamiltonian $H_0[\phi_{\textsc l}]
+H_0[\phi_{\textsc r}]$ as in Eq.~(\ref{HBSG}). Adding an impurity, specifically a local
backscattering term, couples $\phi_{\textsc l}$ and $\phi_{\textsc r}$ at $x=0$. The system
is described by the local sine-Gordon model Eq.~(\ref{HBSG}) with $\beta=\sqrt{8\pi K}$ and 
$\phi$  essentially being the difference of  the bosonic modes on the left and on the right
of the impurity \cite{Fendley95}. We note in passing that for $K< 1/4$ the low-energy mapping
of the Tomonaga-Luttinger model with local impurity onto the local sine-Gordon model
Eq.~(\ref{HBSG}) is expected to break down \cite{Fendley95}.
In this case additional allowed and relevant tunneling
processes across the impurity corresponding to, e.g., the simultaneous tunneling of two
fermions, lead  to higher harmonics like $\cos\left[\sqrt{32\pi K}\phi(0)\right]$.
Accordingly, universality is lost in this case.   

From the scaling dimension $\Delta=K$ of $H_\B$ we recover the RG flow discussed
in section \ref{subsec:lincond}. In the repulsive regime $\Delta<1$, $H_\B$ is a relevant
perturbation that cuts the system and at low energy one reaches the insulating 
fixed-point with vanishing linear conductance. Setting $\gamma=0$ 
in  Eq.~(\ref{HBSG}) corresponds to the perfect conduction fixed point where the
conductance reaches its maximal value $G_0$.

The strong impurity formulation, or cut chain limit, also bears a universal 
description in terms of a (dual, see Refs.~\cite{Kane92,Fendley95}) local sine-Gordon model
$ \widetilde H=H_0+\widetilde H_\B$ with 
$ \widetilde H_\B = \widetilde\gamma\cos\left[\sqrt{\frac{8\pi} K}\phi(0)\right]$.
Setting $\widetilde\gamma=0$  corresponds  to
the cut chain fixed point, i.e. two semi-infinite Tomonaga-Luttinger models.
$\widetilde H_\B$  describes the tunneling of fermions between the two half chains,
and has scaling dimension $\tilde \Delta=K^{-1}$.  Fermion tunneling is therefore
relevant for attractive interactions  $K>1$. 

A bias voltage $V$ applied across the impurity is described by a term 
\begin{equation}
H_V=-eV\sqrt{\frac K{2\pi}}\, \int_{- \infty}^{\infty} \partial_x\phi,
\label{HV}
\end{equation}
which has to be added to the Hamiltonian Eq.~(\ref{HBSG}).
The voltage couples to the charge difference $Q$ between the right and the left of the
impurity. The prefactor ensures that $H_V = - QV/2$.
In the absence of the impurity this leads to the conductance
\begin{equation}
G_0 =K\frac{e^2}h.
\label{defG0}
\end{equation}
Note that we add the fundamental constants such as $e$, $h$,\ldots to the
``theorists units'' whenever we judge this to be appropriate.

The potential term in Eq.~(\ref{HBSG}) generates a typical energy scale,
the ``impurity temperature''
$T_\B$, that scales as $T_\B\sim W^{\frac{\Delta}{1-\Delta}}\gamma^{\frac 1{1-\Delta}}$,
with the scaling dimension $\Delta=K$. Here $W$  denotes the high-energy cutoff
of the model which is left implicit in Eq.~(\ref{HBSG}) and $T_\B$ is the model
specific realization of the nonuniversal temperature $T_0$ introduced in
Sect.~\ref{subsec:lincond}. Low-energy universality implies that physical
quantities $X$ can be expressed as universal functions $f_{\rm univ}^{(X)}$ of
dimensionless arguments $\frac V{T_\B},\frac T{T_\B},\ldots$,
e.g., for the electrical current 
\begin{equation}
I(V,T)=T_\B\,f_{\rm{univ}}^{(I)} \bigg(\frac{V}{T_\B},\frac{T}{T_\B}\bigg) ,
\label{defI}
\end{equation}
as long as $V \ll W$, $T \ll W$, \ldots .  

The local sine-Gordon model is integrable \cite{Ghoshal94}, which allows for a number of exact
predictions for the out-of-equilibrium current Eq.~(\ref{defI}) (for arbitrary 
$\frac V{T_\B}$ and $\frac{T}{T_\B}$) for integer values of $\lambda$ \cite{Fendley95} and
all rational values of $\lambda$ \cite{Boulat19}.
This provides access to the finite temperature linear conductance.
Integrability of the local sine-Gordon model \cite{Ghoshal94,FendleySaleur94,FendleySaleurWarner}
and also the bulk sine-Gordon model,
implies that both possess a rich mathematical structure, with, amongst other things, an infinity
of commuting conserved quantities. Being one of the simplest integrable model, 
the bulk sine-Gordon model has been studied extensively and the Bethe ansatz approach
that will be outlined shortly, builds on developments of formal aspects
that have been studied in great detail. A nonexhaustive list  of significant  contributions
include the construction of quasiparticle modes \cite{Zamolodchikov79,Faddeev80},
and the development of the thermodynamic Bethe ansatz \cite{Zamolodchikov90,Zamolodchikov91}.
In the context of the XXZ spin lattice model, strings and the algebraic Bethe
ansatz were introduced \cite{Takahashi72,TakahashiBook,KorepinBook,FaddeevHouches} and adapted 
to the sine-Gordon model \cite{Fowler82,Chung83,Fendley92,Tateo95}. The  description of the impurity
scattering was included in Refs.~\cite{Ghoshal94,FendleySaleur94,FendleySaleurWarner}.

Avoiding all tedious technicalities that come along with this rich mathematical structure,
in practice one achieves an exact change of the many-body basis from a free
chiral boson with bias voltage (states incoming from $x=-\infty$ towards the impurity),
to a gas of interacting quasiparticles whose thermodynamics is known exactly.
Just as in free gases, where many-body states have the structure of  Fock states
generated by modes $a^\dagger_p$, creating single-particle plane waves with momentum $p$,
the many-body states can be described as collections of quasiparticles created by modes
$A_a(\theta)$, with quantum number $a$ and momentum $p$ parameterized by a rapidity $\theta=\ln p$.
Yet they are interacting quasiparticles, with nontrivial scattering amongst them, which is
encoded in the bulk scattering matrix $S$. As a consequence the density $\rho_a(\theta)$
describing the thermodynamics of the gas depends in a nonlinear way on the densities of
all other particles $\rho_b(\theta')$. The out-of-equilibrium treatment employs integrability
and identifies the basis of quasiparticle modes $A_a(\theta)$ with the following properties:
\begin{enumerate} 
\item \label{ptBos} 
  The many-body states generated by the quasiparticle modes are the many-body states of a
  chiral right moving boson.
\item \label{ptTherm} 
 The density matrix of the free boson gas at $T$ and $V$ can
 be represented in terms of the quasiparticle modes.
\item \label{ptBulk} 
The many-body bulk scattering amongst quasiparticles is factorized.
\item \label{ptDiag} 
The many-body bulk scattering amongst quasiparticles is diagonal.
\item \label{ptImp} 
  The many-body impurity scattering is factorized and without particle production,
  i.e. a single quasiparticle incoming state
  yields a single quasiparticle  outgoing state after impurity scattering.
\end{enumerate}

Finding the basis turns out to be quite different when $\lambda$ is an integer, and 
when it is a positive rational number.

\noindent\underline{Diagonal case}: $\lambda \in \mathbb{N}$\\
The solution to the ``integer" local sine-Gordon model \cite{Fendley95}, with $\lambda\in
\mathbb{N}$, involves two quasiparticles,  a soliton and an antisoliton $A_\pm(\theta)$, carrying an
electric charge $\pm q_{\rm s}=\pm\sqrt{\frac{8\pi K}{\beta^2}}$ respectively, together with 
a collection of $n_{\rm b}=\lambda-1$  neutral bound states $(A_+,A_-)$ called breathers.

\noindent\underline{Off-diagonal case}: $\lambda \in \mathbb{Q}^+$\\
The only recently achieved solution of the ``fractional'' local sine-Gordon model \cite{Boulat19},
involves $N$ quasiparticles with a complex spectrum, whose structure depends on
arithmetic properties of the rational sine-Gordon parameter $\lambda$. In the case $\lambda<1$,
and introducing the integers $\kappa_i$ of the continued fraction
decomposition $\lambda=1/\left[\kappa_1+1/\left\{\kappa_2+... (+1/\kappa_\alpha)\right\}\right]$, the number of modes is 
$N=1+\sum_{i=1}^\alpha\kappa_i$. The first quasiparticle is a neutral soliton $A_{\rm s}$
carrying energy. There are two charged quasiparticles $A_{\rm c}^{\pm}$, with charge $\pm q\times q_{\rm s}$
where $q$ is the denominator of  $\lambda$, and a collection of $N-2$ additional neutral particles,
carrying only entropy, that are also necessary for the proper description.

Once the correct quasiparticles have been identified, a closed formula for the linear conductance
of the  Landauer-B\"uttiker type can be obtained. We here present the  general formula \cite{Boulat19}
valid for the off-diagonal case $\lambda<1$
\begin{equation}
 G(T)=G_0\;{\cal A}_{\lambda} \int_{-\infty}^\infty d\theta\; {\cal T}_\B(\theta)\;\left[-\partial_\theta f_c(\theta)\right]
, \label{condExact}
 \end{equation}  
 where ${\cal A}_\lambda$ is a known numerical constant, and ${\cal T_\B(\theta)}$ is the transmission
 probability of a charged quasiparticle across the impurity, that is also known. The Fermi factor
 $f_c(\theta)=\frac{1}{1+e^{\epsilon_c}}$ is expressed in terms of pseudo energies $\epsilon_a(\theta)$ ($a=1,\ldots,N$)
 that entirely determine the thermodynamical properties
 of the gas. They are in turn determined by a set of nonlinear coupled integral equations, the thermodynamic Bethe
 ansatz equations, 
 \begin{equation}
\epsilon_a(\theta) = \delta_{a,s}\,e^\theta - \frac1{2\pi}\,\sum_{b=1}^N\,K_{ab}\star 
\ln\Big(1+e^{\mu_a-\epsilon_a
}\Big) ,
 \label{pseudoExact}
 \end{equation}  
 where $\star$ denotes a convolution in rapidity space. The kernel $K_{ab}(\theta)$ is known exactly
 and encodes the effect of the interaction among the quasiparticles. The chemical potential reads
 $\mu_a=q\frac{V}{2T}(\delta_{a,N\!+\!1}-\delta_{a,{N}})$.

\begin{figure}
  \includegraphics[width=0.49\textwidth,clip]{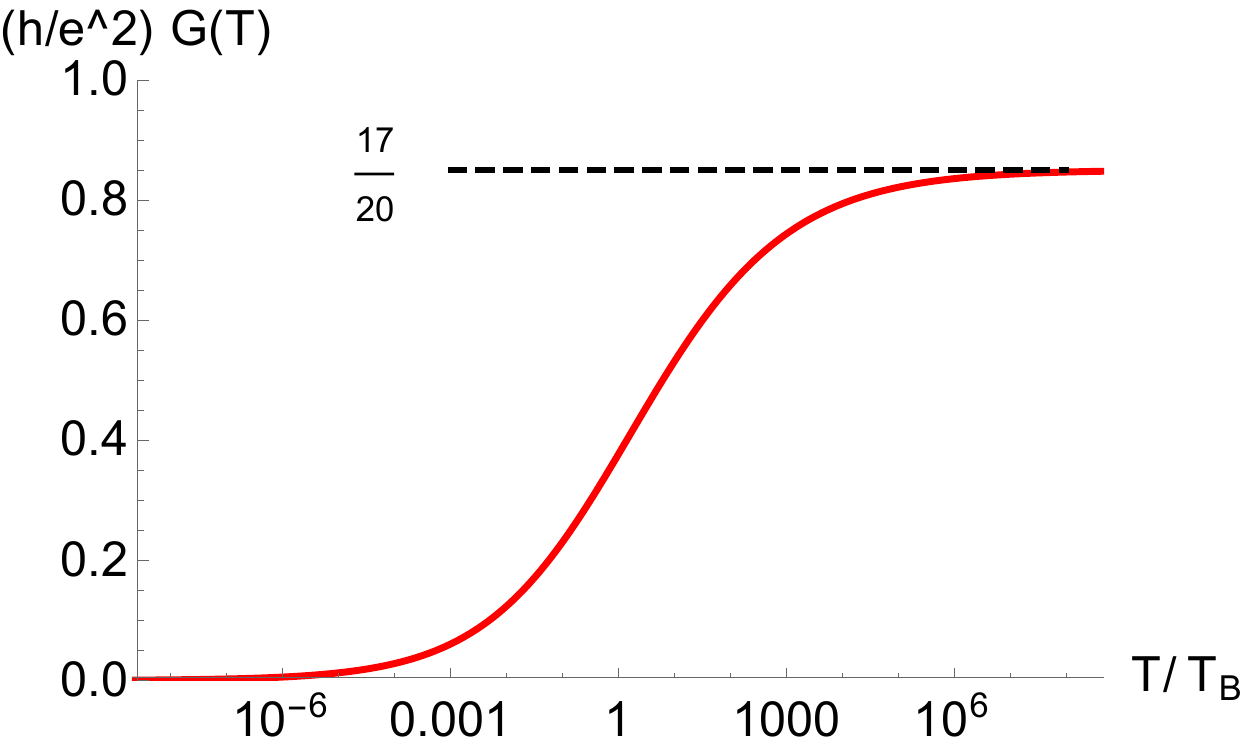}
   \includegraphics[width=0.44\textwidth,clip]{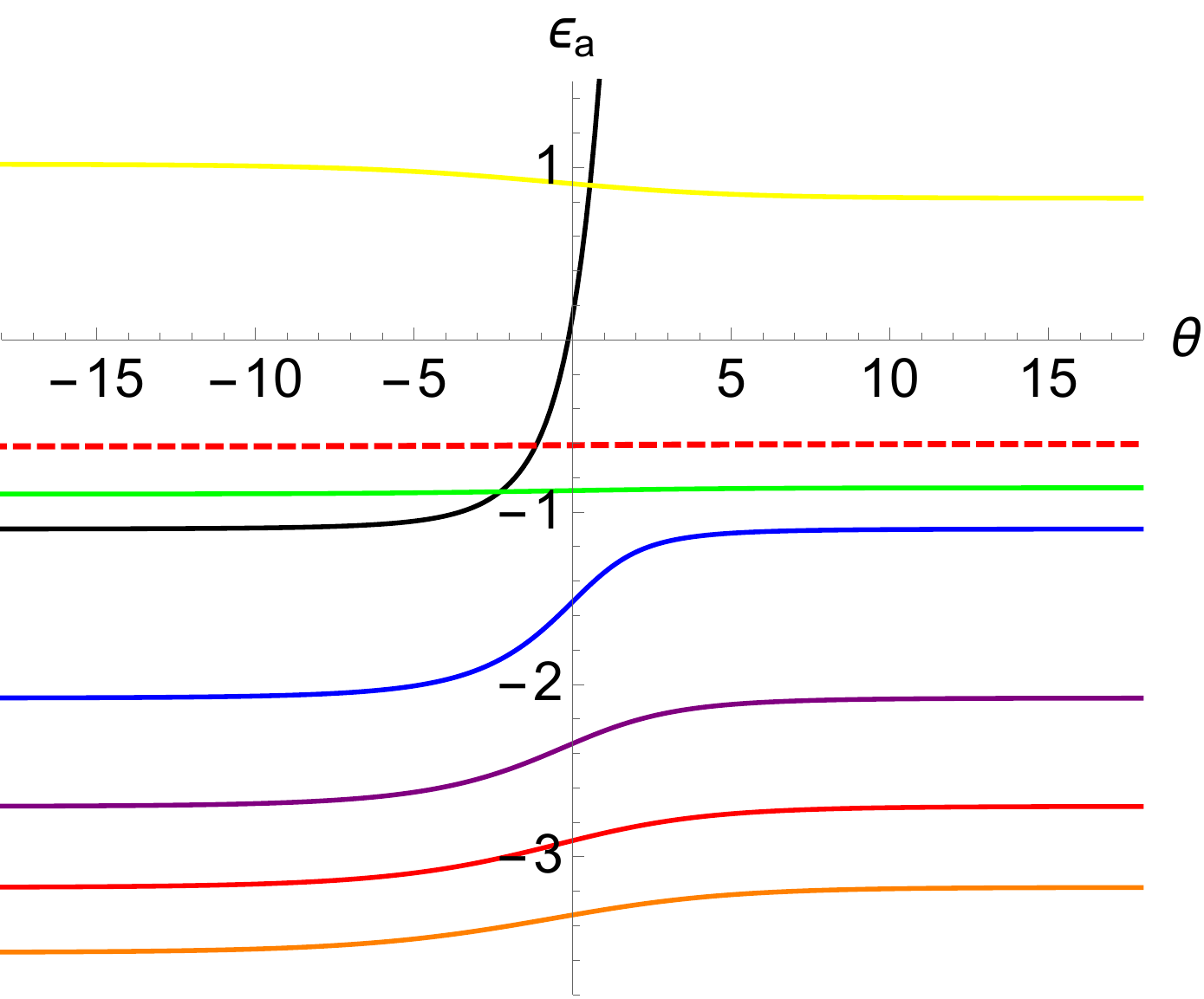}
      \includegraphics[width=0.43\textwidth,clip]{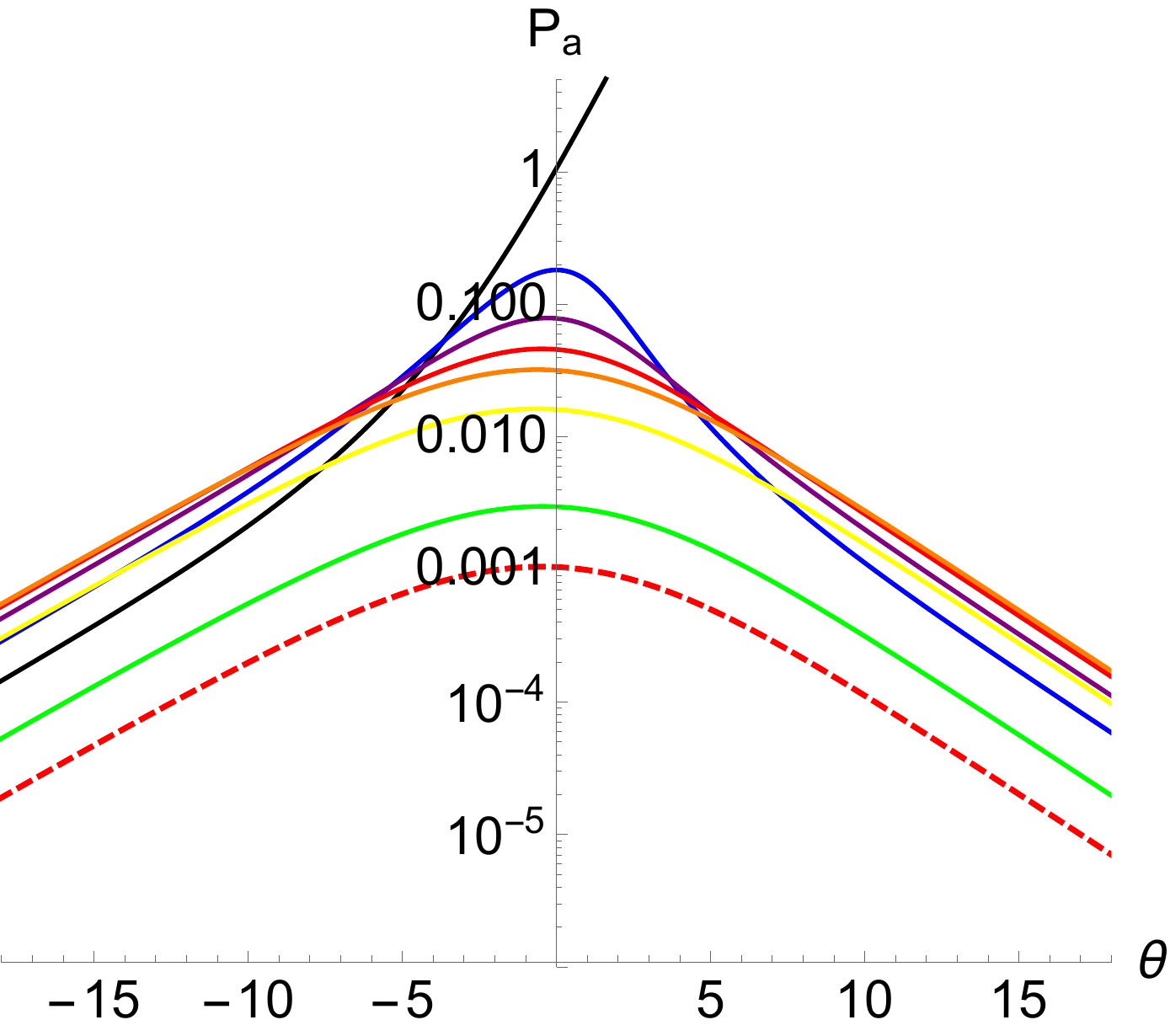}
            \includegraphics[width=0.52\textwidth,clip]{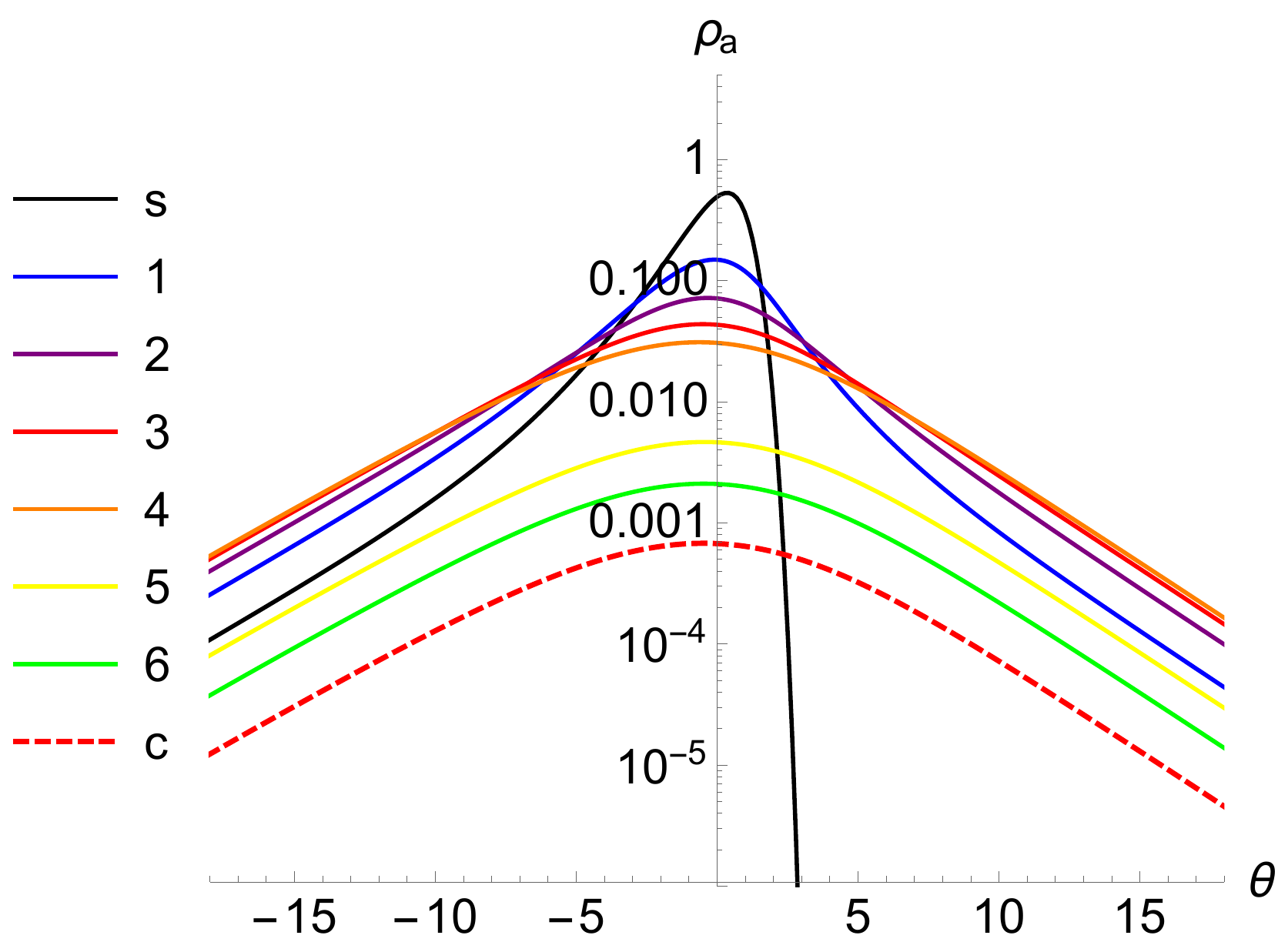}
            \caption{Top left panel: Exact universal linear conductance $G_0 G_K(T/T_\B)$
of the local sine-Gordon model
              for $\lambda=\frac3{17}$
  or $K=\frac{17}{20}=0.85$, obtained from Eqs.~(\ref{condExact}) and (\ref{pseudoExact}).
  Top right panel:  $\theta$-dependence of the pseudo-energies $\epsilon_a$ of the nine quasiparticles
  that are required to describe the system, shown  here in the limit $\frac V T=0$. The two charged
  quasiparticle have degenerate spectrum (dashed line).
  Bottom panel: The corresponding total densities $P_a$ of quasiparticles (left) and the densities
  $\rho_a=P_a f_a$ of occupied quasiparticles (right).}
  \label{fig:TBA17over20}
\end{figure}

We use the case $K=\frac{17}{20}=0.85$ or $\lambda=\frac{3}{17}=1/\left[5+1/(1+1/2)\right]$
involving $1+5+1+2=9$ quasiparticles, to illustrate the procedure for obtaining the conductance.
By numerical integration of the thermodynamic Bethe ansatz equations (\ref{pseudoExact}) one obtains the
pseudo-energies, leading to the conductance $G(T)$ Eq. (\ref{condExact}), see Fig.~\ref{fig:TBA17over20}.
The same procedure but for $\lambda = \frac 13 $ and $ \frac 14$ leads to the dashed lines
in Fig.~\ref{fig:Data-fRG-Sol}.

\section{Spinless fermions on a lattice}
\label{sec:lattice_fermions}

The fermionic many-body model we consider is given by the Hamiltonian
\begin{eqnarray}
H_{\rm hom}  = - t \sum_{j} \big( \,
 c^{\dag}_{j+1} c_j^{\phantom\dag} + c^{\dag}_j \, c_{j+1}^{\phantom\dag}
 \, \big) +  U \sum_{j}  \left( n_j - 1/2 
         \right)  \left( n_{j+1} - 1/2  \right) .
\label{eq:hdef}
\end{eqnarray}
We used standard second-quantized notation with $c^{\dag}_j$ 
and $c_j^{\phantom\dag}$ being creation and annihilation operators on site $j$
respectively and the local density operator $n_j = c^{\dag}_j \,
c_j^{\phantom\dag}$. The nearest-neighbor hopping amplitude is denoted by $t$ and the two-particle
interaction by $U$. We shifted the local density in the interacting part by $1/2$ such that
the Hamiltonian is particle-hole symmetric at half filling $\nu=1/2$. The lattice constant
is set to $a=1$.

In the thermodynamic limit, if the sums over $j$ in Eq.~(\ref{eq:hdef}) run from
$-\infty$ to $\infty$ the homogeneous model can be solved by the Bethe ansatz from which the
Tomonaga-Luttinger liquid parameter $K(U/t,\nu)$ can be extracted (this also
holds for the renormalized velocity $v$, which we, however, are not interested in) \cite{Haldane80}.
At half filling, which we mainly consider, it is given by
\begin{equation}
 K = \left[\frac{2}{\pi} \, 
 \arccos \left(-\frac{U}{2 t} \right) \right]^{-1}.
\label{eq:BetheAnsatz}
\end{equation}
The model belongs to the Tomonaga-Luttinger liquid universality class for $-2 < U/t < 2$.
For other fillings $K(U/t,\nu)$ can be computed by numerically solving a set of integral
equations and the model is a Tomonaga-Luttinger liquid for all $U/t>-2$ \cite{Haldane80}.
For later use we report the expansion of $K$ in $U/t$ for arbitrary filling \cite{Giamarchi03}
\begin{equation}
  K=1- \frac{U}{\pi v_{\rm F}} \left[ 1 - \cos \left( 2 k_{\rm F} \right) \right] +
  {\mathcal O}\left( [U/t]^2
    \right) ,
\label{eq:BetheAnsatzexpand}
  \end{equation}
with the Fermi velocity $v_{\rm F} = 2 t \sin k_{\rm F}$ and $k_{\rm F}= \nu \pi$ . 

The formalism we use below to compute the temperature dependence of the linear
conductance of the model Eq.~(\ref{eq:hdef}) complemented by a local
impurity, employs a transport geometry in which the interacting wire is connected
to two semi-infinite noninteracting leads. We restrict the interaction to $N-1$ bonds
between the sites $j \in [1,N]$ and in the second sum in Eq.~(\ref{eq:hdef}) the site index
thus runs from $j=1$ to $j=N-1$. However, the lattice site in the first sum runs from
$-\infty$ to $\infty$. The sites $j \leq 0$ form a left lead and the sites $j \geq N+1$ a right
one. Due to the abrupt change of the two-particle interaction at the two contacts at
sites $j=1$ and $j=N$ the $T=0$ conductance of the setup does not take the unitary
value. The inhomogeneity of the two-particle
term leads to a single-particle backscattering which masks the effect of
the single impurity to be introduced into the interacting part of the wire. We therefore
smoothly turn off the interaction over $N_{\rm c} \ll N$ lattice sites when approaching the
contacts at $j=1$ and $j=N$ from the center on the interacting chain. As discussed in
Ref.~\cite{Janzen06} the details of the envelope function do not matter as long as it is
sufficiently smooth. This way the $T=0$ conductance in the absence of an impurity
can be tuned arbitrarily close to
the unitary limit. In the numerical results shown below, me made sure that the relative
deviation from the unitary conductance is less than $10^{-4}$.
Note that for this setup with ``adiabatically connected''
noninteracting leads the unitary conductance is given by
$G_0=e^2/h$, instead of $K e^2/h$ as obtained
for the model of Sect.~\ref{sec:sine_Gordon} in which the interaction is not restricted
to a subsystem \cite{Janzen06,Safi95,Maslov95}. One can expect that this difference
does not affect the universal properties after adding a local impurity.
Below we will confirm this by directly comparing the conductance and the $\beta$-function
of the present model with the one of the local sine-Gordon model. 

The system is complemented by a hopping impurity on the bond from site $N/2$ to site
$N/2+1$ ($N$ even) with the Hamiltonian
\begin{equation}
  H_{\rm hop}  = t' \left(
  c^{\dag}_{N/2+1} c_{N/2}^{\phantom\dag} + c^{\dag}_{N/2} \, c_{N/2+1}^{\phantom\dag} \right) ,
\label{eq:hopimphamdef}
\end{equation}
or a site impurity with $H_{\rm site}=V n_{(N+1)/2}$, $N$ odd.  
In the first case the total hopping across the central bond is
$t-t'$. A $t' \neq 0$ or $V \neq 0$ leads to a nonvanishing impurity backscattering.
For $U=0$ the transmission amplitude and thus
$G(T)$ for both types of impurities can be computed exactly using single-particle
scattering theory (see, e.g. Ref.~\cite{Meden98}).

To compute $G(T)$ of the lattice model for $U \neq 0$ we use Matsubara Green functions
and the functional RG approach \cite{Metzner12}.
The technical details of the application of this method to study transport properties of lattice
models of inhomogeneous Tomonaga-Luttinger liquids are given in Refs.~\cite{Meden04,Enss05,Andergassen06}. We here only present the basic idea. The relevant steps are the following:
\begin{enumerate}
\item Express the partition function as a coherent state functional integral.
\item Integrate out the noninteracting leads by
    projection. They are incorporated exactly as lead self-energies in the propagator of the
    interacting part.
\item Replace the reservoir-dressed noninteracting propagator of the system by one decorated by
    a cutoff $\Lambda$. For the initial value $\Lambda_{\rm i}$, the free propagation
    must vanish; for the final one $\Lambda_{\rm f}$, the original propagation must be restored.
    One often uses a cutoff in the Matsubara frequency. When $\Lambda$ is sent from $\infty$ to
    $0$ this incorporates the RG idea of a successive treatment of energy scales. 
\item Differentiate the generating functional of one-particle irreducible vertex functions
    with respect to $\Lambda$.
\item Expand both sides of the functional differential equation with respect to the one-particle
    irreducible vertex functions. This leads to an infinite hierarchy of coupled differential
    equations for the vertex functions. 
\end{enumerate} 
The hierarchy of coupled flow equations presents an exact reformulation of the quantum many-body problem.
Integrating it from $\Lambda_{\rm i}$ to $\Lambda_{\rm f}$ leads to exact expressions
for the vertex functions from which observables such as the conductance can be
computed. In practice, truncations of the hierarchy are required, resulting in
a closed finite set of equations. The integration of these leads to approximate expressions
for the vertices and, thus, for observables.

Truncating the infinite hierarchy of equations we neglect the flow of three- and higher particle vertex
functions and replace the two-particle vertex by a static one of
nearest-neighbor-type. The resulting flow equations and the expression of $G(T)$
in terms of the vertex functions can be found in Refs.~\cite{Meden04,Enss05}.
In this truncation the self-energy (single-particle vertex function) is independent
of the Matsubara frequency and can be interpreted as an effective single-particle
potential which is generated during the RG flow by the interplay of the bare impurity
and the two-particle interaction. At the end of the flow (for $\Lambda_{\rm f}$) a
single-particle scattering problem in the presence of the effective potential has to be solved.
This provides a comparatively simple picture of a many-particle correlation effect.
Integrating the coupled flow equations for the effective interaction and the
self-energy in the limits $t' \to 0$ or $V \to 0$
(weak impurity) and $t' \to t$ or $V \to \infty$ (weak link) it was shown analytically that the
approximate functional RG reproduces the power-law scaling underlying Eqs.~(\ref{eq:Gweakimp})
and (\ref{eq:Gweaklink}) \cite{Meden02}.
In the effective single-particle picture the power laws follow from scattering off
an effective potential which during the RG flow develops a long-ranged oscillatory
part \cite{Meden04,Enss05} (see also Ref.~\cite{Yue94})  out of the purely local bare impurities
(hopping or site). Further away from the bare impurity the long-range
part leading to the power-law scaling
of $G(T)$ is cut off at a scale $\sim 1/T$.

\begin{figure}
  \includegraphics[width=0.48\textwidth,clip]{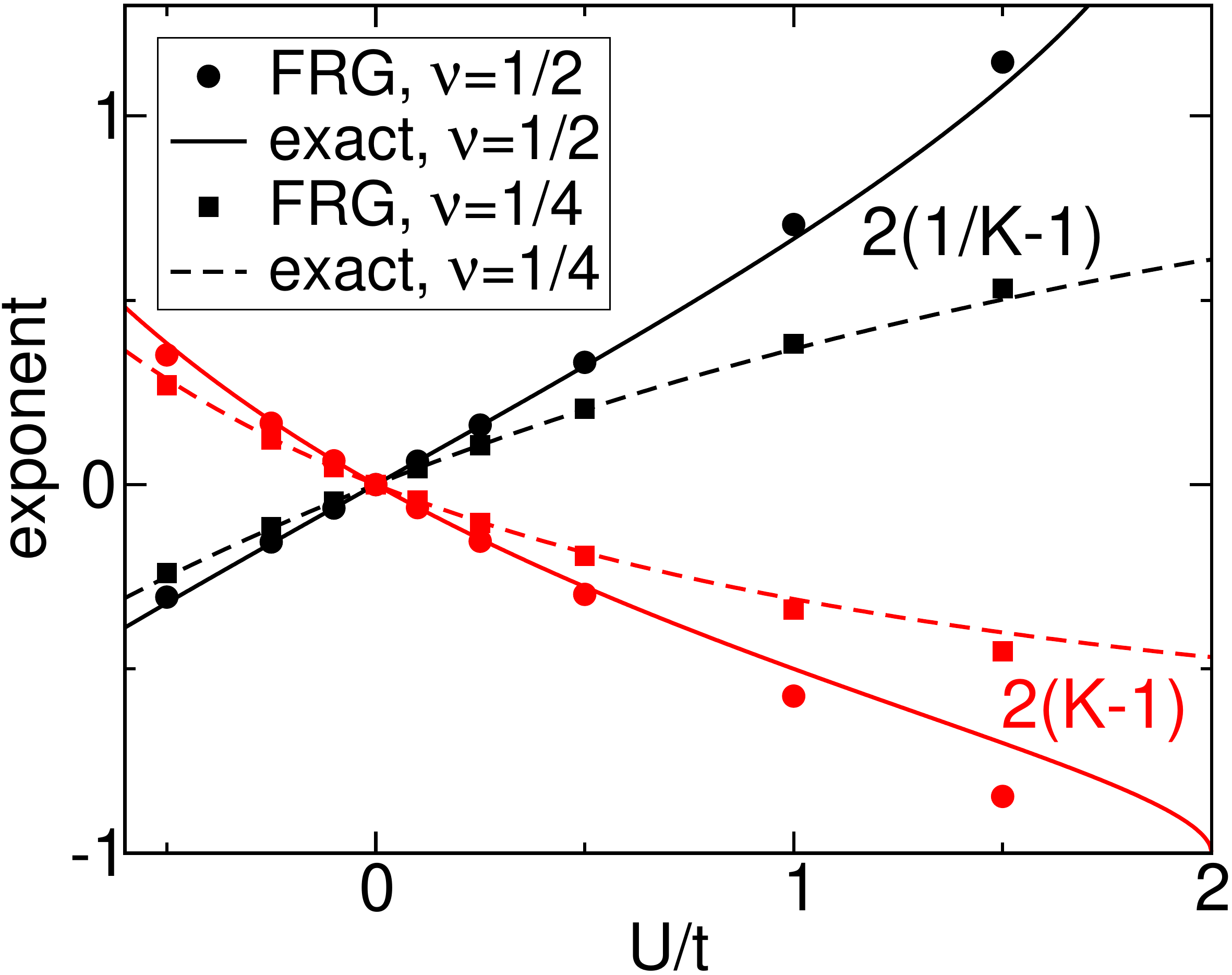}
   \includegraphics[width=0.48\textwidth,clip]{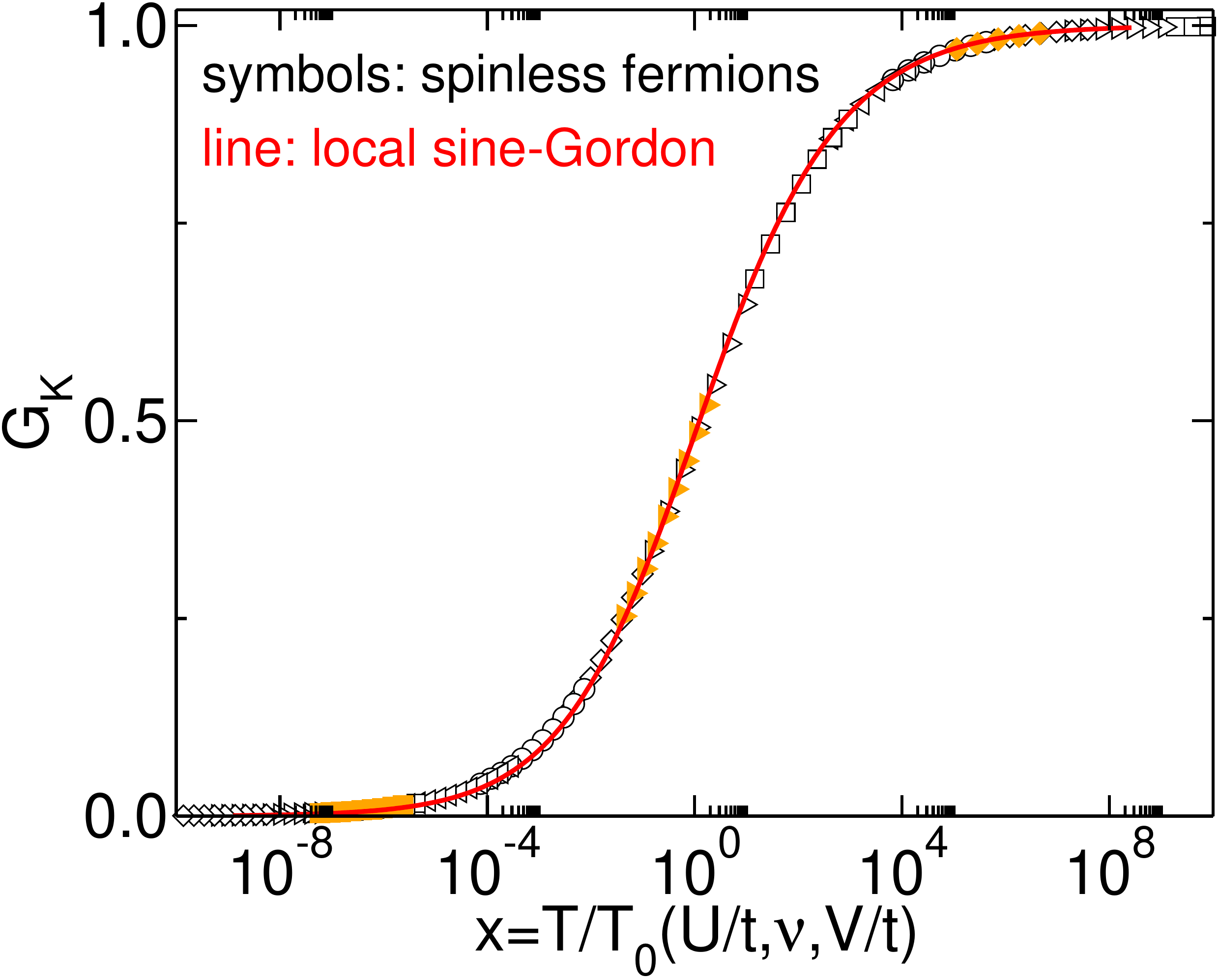}
  \caption{Left panel: Strong (black) and weak (red) impurity exponents as a function of $U/t$
    for two different fillings $\nu=1/2$ and $1/4$. Right panel:
    One-parameter scaling plot of the conductance. Open
  symbols represent results obtained for $U/t=0.5$, $\nu=1/2$, and
  different $T$ and amplitudes of the site impurity $V$,
  while filled symbols were calculated for 
  $U/t=0.851$, $\nu=1/4$. Both pairs of $U/t$ and $\nu$ lead to the same
  $K=0.85$. The solid lines
  shows the universal conductance curve of the local sine-Gordon model for $K=0.85$
  (see the upper left panel of Fig.~\ref{fig:TBA17over20}).}
  \label{fig:FRG}     
\end{figure}

In the left panel of Fig.~\ref{fig:FRG} we show the exponents $2(K-1)$ and $2(1/K-1)$ as functions of
$U/t$ for two different fillings, obtained by inserting the Bethe ansatz result for $K$
(``exact'') as well as by numerically computing $G(T)$ from the functional RG and extracting
the exponent (``FRG''). 
For sufficiently small $|U/t|$ the agreement is excellent, and the deviations from the exact results are
small up to moderate values of the interaction. In fact, the extracted exponents agree to leading order in $U/t$
to the result obtained when inserting the expansion Eq.~(\ref{eq:BetheAnsatzexpand}) for $K$.

The right panel of Fig.~\ref{fig:FRG} shows a one-parameter scaling
plot of the conductance in the presence of a site impurity as a function of
$x=T/T_0(U/t,\nu,V/t)$  with a nonuniversal scale
$T_0(U/t,\nu,V/t)$ \cite{Kane92,Moon93,Leung95,Fendley95,Meden03,Enss05}.
Open symbols were computed for $\nu=1/2$ and $U/t=0.5$, while filled ones were computed for
$\nu=1/4$ and $U/t=0.851$. For both parameter sets one obtains $K \approx 0.85$ (within the approximated functional RG).      
For appropriately chosen $T_0$ (determined ``by hand'') the $G(T)/G_0$ curves for different $V$, $\nu$, and $U$, 
but fixed $K$, collapse onto a $K$-dependent, dimensionless scaling function $G_K(x)$.
Note that the data extend over roughly 20 orders of magnitude in temperature. 
In accordance with Eqs.~(\ref{eq:Gweakimp}) and (\ref{eq:Gweaklink})  $G_K(x)$ exhibits the limiting
behavior $G_K(x) \propto 1-x^{2(K-1)}$ for $x \to \infty$ and $G_K(x) \propto x^{2(1/K-1)}$
for $x \to 0$. We note that to prevent deviations from scaling $T$ must be taken much
smaller than the band width $4 t$---the high-energy cutoff of the present model---and much larger
than a scale of order $1/N$ which cuts off the universal behavior at the lower end. In the figure $N=10^4$
was considered such that, for fixed parameters, several orders of magnitude in $T$ can be used
for the scaling plot. We note that in addition data for a hopping impurity can be collapsed
on the curve (not shown). For fixed $K$ this provides a ``numerical proof'' of universality,
that is independence on the details of the impurity and its amplitude, the interaction strength
$U/t$ as well as the filling $\nu$, within the lattice model. For the local sine-Gordon model
this type of ``internal'' universality can be shown analytically; see Sect.~\ref{sec:sine_Gordon}.  

In addition to the functional RG data for the spinless fermion model (symbols) the right panel
of Fig.~\ref{fig:FRG} shows the universal linear conductance (divided by $G_0 = K e^2/h$) of the local
sine-Gordon model (red line; the same data as in the upper left panel of Fig.~\ref{fig:TBA17over20}).
The agreement is excellent. It is plausible to assume that the small deviations result out of the
approximate nature of our functional RG approach to the spinless fermion model which is only controlled
for small to intermediate $|U/t|$. Decreasing $|U/t|$, i.e.~bringing $K$ closer to one, further
reduces the deviation. This indicates that the low-energy physics of both models, with $T \ll W$ for the
local sine-Gordon model and $v_{\rm F}/N \ll T \ll 4 t$ for the lattice model of spinless fermions with
an impurity, is dominated by the same quantum critical point. The excellent
agreement ``proves''
universality across the models. We here compare the $G_K(x)$ of the two models, and not the
$\beta$-function, as it has a direct physical meaning (the conductance).     

After introducing the experimental
quantum circuit as our third system to realize the quantum critical point in the next section,
we will show that the $\beta$-function,
in which the nonuniversal scale is eliminated, is the same in all cases and, therefore, universality
holds between all three systems. This will also confirm our expectation, that $\beta_K$ is a function
of $g$ only, see Sect.~\ref{subsec:lincond}.  

\section{A tunable quantum circuit: experiments}
\label{sec:quantum_circuit}

\begin{figure}
\center
  \includegraphics[width=60mm]{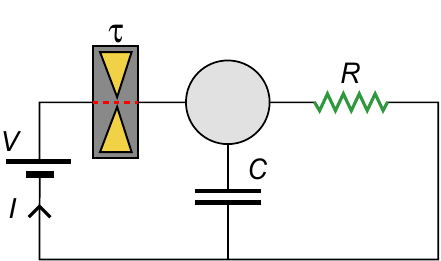}
  \caption{Schematic representation of a quantum circuit described at low temperature
    ($k_\mathrm{B}T\ll h/RC$) by the same Hamiltonian as the local sine-Gordon model and
    the model 1d spinless fermions with an impurity. In practice, a series resistance
    $R=h/ne^2$ is implemented from the parallel combination of $n$ quantum Hall edge channels.
  }
  \label{fig:CircuitSchem}     
\end{figure}

A quantum circuit composed of a short spin-polarized electronic channel in series with a resistance
(see schematic in Fig.~\ref{fig:CircuitSchem}) appears markedly different from a system of interacting
electrons confined to 1d. Nevertheless, at low temperatures, this circuit is predicted to be
mathematically described by the same local sine-Gordon model \cite{Safi04} as a 1d metal of
spinless fermions including a single impurity \cite{Giamarchi03}. 
Such a connection illustrates the remarkable universal character of the Tomonaga-Luttinger liquid concept.
It originates from the similar continua of bosons describing, on the one hand, 1d fermions with short-range
interactions \cite{Haldane80,Giamarchi03} and, on the other hand, a linear resistance $R$ in a quantum
circuit \cite{SCT92}. As a result, the two different systems can be mapped onto the same Tomonaga-Luttinger Hamiltonian
with interaction parameter $K=1/(1+Re^2/h)$ \cite{Safi04}. In practice, the quantum circuit
implementation \cite{Parmentier11,Mebrahtu12,Jezouin13,Anthore18} constitutes an experimental test-bed that
stands out in that it allows for direct, quantitative and parameter-free investigations of the universal
Tomonaga-Luttinger liquid physics; see our discussion in Sect.~\ref{subsec:homTL} for other realization
of Tomonaga-Luttinger liquids.

Quantum circuits can be engineered or adjusted in-situ to cover arbitrary strengths of repulsive interactions
$0<K<1$, through the tuning and separate characterization of
$R$ \cite{Parmentier11,Mebrahtu12,Jezouin13,Anthore18}.
For the data discussed in this minireview, we used a robust and precise approach that consists in implementing
a linear resistance $R=h/ne^2$ from $n\in\mathbb{N}$ integer quantum Hall edge channels in
parallel \cite{Jezouin13,Anthore18}. Correspondingly, the interaction parameter can take the values $K=n/(n+1)$,
with here $n\in\{1,2,3,4\}$. Note that lower values of $K$ can be achieved through larger resistances
obtained with narrow metallic stripes (instead of ballistic channels) \cite{Parmentier11}.
On a practical side, such stripes should be short enough so that any distributed capacitance remains negligible
with respect to $R$ up to the relevant frequency scale $k_\mathrm{B}T/h$, with $k_\mathrm{B}$ the Boltzmann constant. 

The impurity is realized by a quantum point contact formed in a Ga(Al)As two dimensional
electron gas, see Fig.~1 of Ref.~\cite{Anthore18} for an electron micrograph of the device schematically represented
in Fig.~\ref{fig:CircuitSchem}. The spin polarization is achieved by immersing the device in a large perpendicular
magnetic field of $2.7\,$T  (corresponding to the regime of the integer quantum effect at filling factor 3). The impurity
quantum point contact is tuned in-situ, by field effect, to partially transmit a single electronic channel (the
outer quantum Hall edge channel). It is characterized by the unrenormalized transmission probability $\tau$ of
electrons across the channel, which is experimentally obtained from the quantum point contact differential
conductance $G_\mathrm{QPC} = \tau e^2/h$ measured at large enough dc voltage bias to suppress the
Tomonaga-Luttinger (dynamical Coulomb blockade) conductance reduction.

The interconnection between the quantum point contact and series resistance $R=h/ne^2$ involves a micron-scale metallic
island playing the role of a floating reservoir merging all quantum Hall channels, which would otherwise separately
propagate along opposite sample edges. Note that the metal is thermally diffused into the GaAs-GaAlAs heterojunction,
in order to make a good electrical contact with the two dimensional electron gas located approximately 100\,nm below
the surface. The geometrical self-capacitance of the island $C\simeq 3.1$\,fF determines the relevant high-energy
cutoff, with universal Tomonaga-Luttinger behavior emerging at $k_\mathrm{B}T\ll h/(RC)$. The knowledge of the important
cutoff parameter $C$ is experimentally obtained from Coulomb diamond measurements of the conductance across the same
island, but with all connected quantum point contacts tuned to weak tunnel couplings \cite{Anthore18}.

This circuit can be described by the following Hamiltonian:
\begin{equation}
H=H_0+H_{\rm I}+H_{\rm C}+H_\mathrm{env}.
\end{equation}
Here $H_0$ represents a ballistic conduction channel
\begin{eqnarray}
H_0=i\hbar v_\mathrm{F}\int dx\,\left(\psi^+_{+}\partial_x\psi_{+}
  -\psi^+_{-}\partial_x\psi_{-}\right) 
\label{eq:H0exp}
\end{eqnarray}
where $\psi_{+(-)}$ is the annihilation operator for the electrons moving
toward (away from) the island and $v_\mathrm{F}$ is the Fermi velocity. $H_{\rm I}$ models the backscattering at the
QPC located at $x=0$, $H_\mathrm{I}=\hbar v_\mathrm{F}r\left[\psi^+_{+}(0)\psi_{-}(0)+\psi^+_{-}(0)\psi_{+}(0)\right]$,
with $|r|^2\simeq1-\tau$ for a near ballistic QPC. $H_{\rm C}$ is the coupling between electrons in the channel and
the electromagnetic RC environment, $H_\mathrm{C}=-\hat{Q}(V-\partial_t\hat{\Phi})$ with $\hat{Q}$ the total charge
transferred across the QPC and $\hat{\Phi}$ a bosonic operator corresponding to the time integral of the voltage
across the RC impedance. The dissipative RC environment $H_\mathrm{env}$ is modeled by an infinite set of quantum
LC resonators as detailed in \cite{SCT92}. The mapping of such a circuit on the spinless Tomonaga-Luttinger Hamiltonian
including an impurity was first demonstrated in \cite{Safi04}.

The prominent observable for electronic systems is the electrical conductance.
In the present context of a Tomonaga-Luttinger liquid with an impurity, a crossover from a conductor toward an insulator
is predicted to develop as the temperature is reduced, except in the absence of interactions $K=1$ ($R=0$) or in
the absence of any impurity ($\tau=1$) \cite{Apel82,Kane92}. According to the (inhomogeneous) Tomonaga-Luttinger
liquid theory, and as shown explicitly in Sect.~\ref{sec:lattice_fermions},
the conductance $G$ along this crossover follows a universal scaling law that only depends on
the interaction parameter $K$ \cite{Kane92}.

\begin{figure}
  \includegraphics[width=\textwidth]{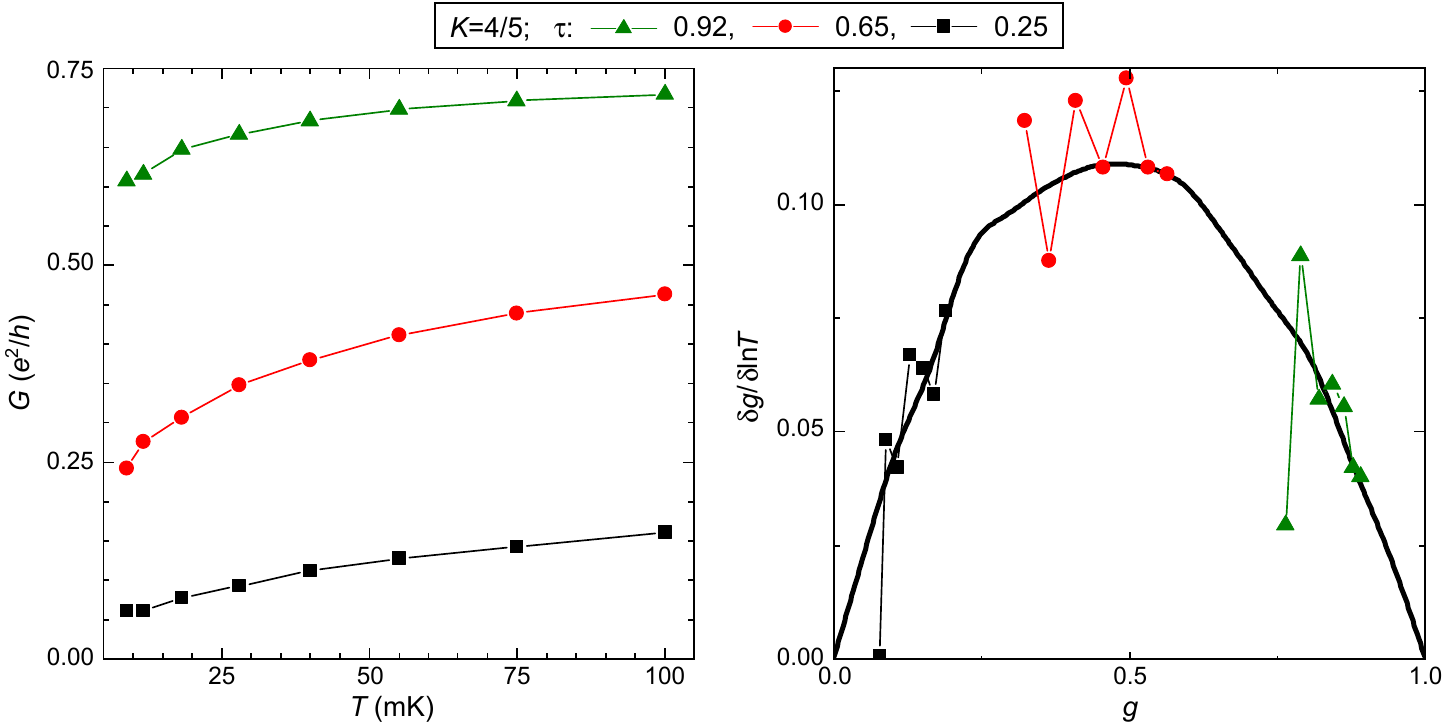}
  \caption{Experimental procedure illustrated at $K=4/5$ ($R=h/4e^2$) for three settings of the impurity backscattering $\tau$.
  Left panel: Symbols represent measurements of the conductance across the device plotted versus temperature. 
  Right panel: Discrete differentiation $\delta g/\delta \ln T$ plotted versus $g  = G/(Ke^2/h)$.
  Symbols connected by lines are obtained from the conductance data of corresponding color in the left panel.
  The thick continuous lines represent a low-pass Fourier averaging of the full data set ($\sim200$ values of $\tau$).
  }
  \label{fig:Expt_G&dg/dlnT}     
\end{figure}

\begin{figure}
\center
  \includegraphics[width=90mm]{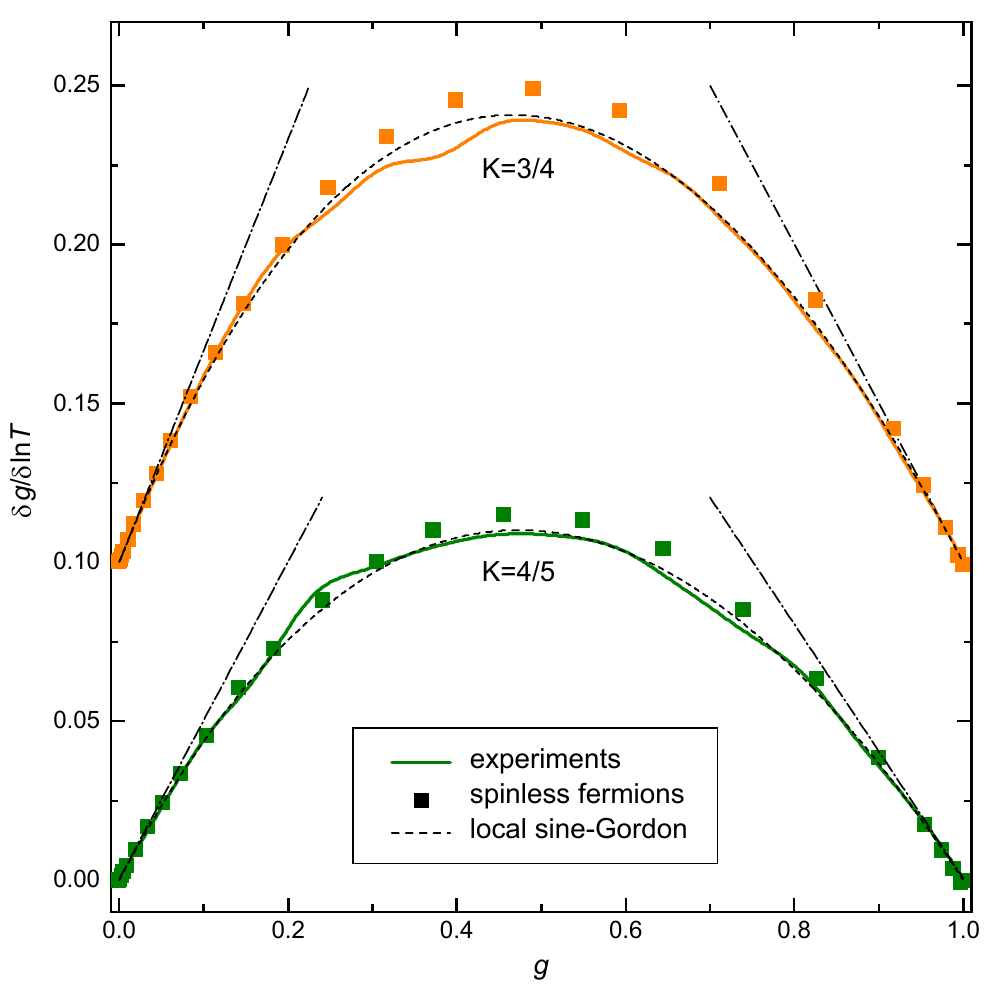}
  \caption{Comparison of the renormalization flow $\beta$-functions at $K=3/4$ (shifted vertically by $0.1$)
    and $4/5$ obtained experimentally by measuring a quantum circuit (colored continuous lines), numerically
    by computing the conductance employing an approximate functional RG approach to a model of interacting
    spinless fermions with one impurity (symbols), and analytically by solving exactly the local sine-Gordon
    model (blacked dashed lines). The asymptotic behavior Eqs.~(\ref{eq:betaweakimp}) and (\ref{eq:betaweaklink})
    for $1-g \ll 1$ and $g \ll 1$, respectively, is indicated by the dashed-dotted lines.
  }
  \label{fig:Data-fRG-Sol}     
\end{figure}

Experimentally, for all implemented interaction parameter $K=1/(1+Re^2/h)\in\{1/2,2/3,3/4,4/5\}$, the impurity
backscattering strength is spanned over the full range $\tau\in[0,1]$. At each given setting of $\tau$, we measure
the conductance $G$ of the device (quantum point contact and series resistance) at several values of the
temperature, as illustrated for $K=4/5$ ($R=h/4e^2$) in the left panel of
Fig.~\ref{fig:Expt_G&dg/dlnT} with $\tau\simeq0.25$ (black), $0.65$ (red) and $0.92$ (green).
Then, using data points up to temperatures of at most $h/(25k_\mathrm{B}RC)$ (in order to ascertain a
universal low energy behavior), we perform discrete differentiations $\delta g/\delta \ln T$ for each given
device setting of $\tau$. The result is plotted as a function of $g$ in the right panel of
Fig.~\ref{fig:Expt_G&dg/dlnT}, as symbols connected by lines corresponding to the conductance data of the same color
in the left panel. This procedure is repeated for many settings of $\tau$, approximately 200 values for each $K$. 
As shown in Fig.~2(a) of Ref.~\cite{Anthore18}, all data points for a given $K$ (series resistance) pile up on the same
curve independently of the device's tuning of $\tau$, which directly demonstrates an underlying universal
scaling behavior. The corresponding experimental renormalization flow $\beta$-function $\beta_K(g)$, shown
as thick continuous lines in the right panel of Fig.~\ref{fig:Expt_G&dg/dlnT} for $K=4/5$ and in Fig.~\ref{fig:Data-fRG-Sol}
for $K=3/4$ and $4/5$, were obtained by a low-pass Fourier averaging of the individual data points
(for $K=1/2$ and $2/3$ see Fig.~2(a) of Ref.~\cite{Anthore18} including a comparison with exact solutions of the
local sine-Gordon model).
The quantitative agreement reached between the experimentally derived $\beta_K(g)$ using a quantum
circuit implementation (colored continuous lines), the exact theoretical solutions of the local sine-Gordon model
(black dashed lines), and the functional RG results for a lattice model of spinless fermions with one
impurity (symbols) is displayed in Fig.~\ref{fig:Data-fRG-Sol}. We emphasize that the comparison is free of any
fitting parameters. As expected (see Sect.~\ref{subsec:lincond}) the unique $\beta$-function depends
on the temperature only via the dimensionless conductance $g$. The parameter free collapse of all data sets   
provides a convincing example of the power of emergent universality originating from an underlying
quantum critical point. This universality provides a bridge between different fields of physics, here between
field theory, quantum many-body theory of correlated Fermi systems, and experimental circuit quantum
electrodynamics.

\section{Summary and outlook}
\label{sec:sum}

There exist at least three levels from which the results presented in our minireview can be summarized.

The highest one is the perspective of emergent universality in complex systems resulting out of a
common underlying (quantum) critical point. It bridges different fields of physics. Our theoretical
results obtained for the field theoretical local sine-Gordon model and the condensed matter
model of spinless lattice fermions with nearest-neighbor hopping, nearest-neighbor interaction
and a local impurity, as well as the experiments on highly tunable quantum electrodynamical
circuits provide a convincing example of the power of this concept. The parameter free collapse
of the low-energy $\beta$-function of the conductance for different $K$, as shown in
Fig.~\ref{fig:Data-fRG-Sol}, is of exceptional high quality. It explicitly shows that the underlying
quantum critical point is the same for all three systems.

The second level is the one of the physics of a local impurity in a 1d correlated Fermi systems. Considering the two
different models and the experimental emulation we were able to shed light on the emergence of the
metal-to-insulator transition. Even if the bulk system is metallic the interplay of the single impurity
and a repulsive two-particle interaction will drive it into an insulating state. The quantum critical
nature of the transition leads to the universality within a given system. For fixed $K$, the microscopic
details and parameters enter the conductance only via a nonuniversal scale $T_0$.

The third level concerns the experimental verification of the (inhomogeneous) Tomonaga-Luttinger liquid concept.
Our discussion shows that to convincingly demonstrate this type of physics requires experimental control,
tunability, and the access to the variable in which power-law scaling is to be shown (the rescaled temperature $T/T_0$ with a temperature scale $T_0$ encapsulating microscopic details)
over several orders of magnitude. In this respect the emulation by quantum circuits is clearly superior to
the attempts to directly realize (quasi-) 1d fermionic systems in semi-conductor-based heterostructures,
self-organized atom chains on surfaces, and unidirectional long molecules (e.g. metallic carbon nanotubes).
We are not aware of any experiments on such systems which show Tomonaga-Luttinger liquid behavior in a way
which is equally convincing as the emulations discussed here (see also Ref.~\cite{Mebrahtu12}). In fact,
many of the interpretations of experimental data on (quasi-) 1d fermionic systems in the light of
Tomonaga-Luttinger liquid theory have been questioned. 

As a next step it would be very interesting to emulate other models of inhomogeneous Tomonaga-Luttinger
liquids. Even introducing a second localized impurity is expected to lead to new effects associated
to resonant transport (see Refs.~\cite{Meden08,Enss05} and references therein). In this case no
exact solution for any model properly describing this situation is available. In fact, different
approximate approaches lead to conflicting results on the emergence of a regime of energies with a
novel scaling exponent. The experimental
emulation of this setup could provide a very useful contribution to settle this issue. Novel fixed
points with unique scaling exponents have also been predicted for Y-junctions of three Tomonaga-Luttinger
liquids (see Ref.~\cite{Meden08} and references therein). Experimentally realizing those
would also be of great interest. Furthermore, experimentally emulating correlated 1d fermions in
(more) complex nonequilibrium situations, such as periodically driven systems, could help to
deepen our understanding of the interplay of two-particle interactions and nonequilibrium.
Currently, the number of tools to investigate such systems theoretically in a controlled way is
very limited.

\end{document}